\documentstyle[preprint,pra,aps]{revtex}
\draft
\begin{document}
\title{Confined Gap Excitations and Multielectron Localization
in Magnetic Alloys}
\author{Valery I. Rupasov\cite{pa}}
\address{Landau Institute for Theoretical Physics,
Russian Academy of Sciences, Moscow, Russia}
\date{\today}
\maketitle
\begin{abstract}
The degenerate Anderson model with a nonlinear electron
dispersion and an energy dependent hybridization is proven
to exhibit hidden integrability and is diagonalized by the
Bethe ansatz. If the impurity $f$-level energy lies below
the conduction band edge, the spectrum of the system is
shown to contain pairs of confined ``gap excitations''
and their heavy bound complexes. In addition, multielectron
localized states due to a pinning of gap bound
complexes to the impurity are predicted.
\end{abstract}


It is well known that the basic theoretical models describing
magnetic impurities in nonmagnetic metals \cite{MA}, such as
the s-d (Kondo) model, the Anderson model, etc., are integrable
under {\em two additional conditions}: (i) an electron-impurity
coupling is assumed to be energy independent, and (ii) a band
electron dispersion $\varepsilon(k)$ can be linearized around
the Fermi level, $\varepsilon(k)\simeq v_F(k-k_F)$, where
$k_F$ and $v_F$ are the Fermi momentum and velocity, respectively.
Only under these conditions, both an electron-impurity scattering
and an effective electron-electron coupling are described in terms
of discontinuous jumps in the Bethe ansatz wave functions. Therefore
a linear particle spectrum and a pointlike particle-impurity coupling
are considered now as the {\em necessary} mathematical conditions
for integrability of the ``impurity'' models. In what follows we will
use the phrase ``linear dispersion approximation'' (LDA) to refer
to a model modified in accordance with these conditions. The LDA
is clear to eliminate both the effects of the band structure and
possible excitations of the system with energies lying inside the
interband gap from an exact analysis of the behaviour of
magnetic alloys. Therefore an extension of the Bethe ansatz technique
to models accounting for a more realistic band structure and an energy
dependence of particle-impurity coupling is of great physical
interest.

In the present paper, we study the $n$-fold degenerate Anderson
model with a nonlinear band electron dispersion and an energy
dependent electron-impurity coupling, which is the basic theoretical
model for mixed-valent, Kondo and heavy-fermion phenomena \cite{MA}.
The model Hamiltonian is written in the following one-dimensional form:
\begin{mathletters}
\begin{equation}
H=\sum_{\alpha}\int_{0}^{\infty}\frac{dk}{2\pi}
\left\{\varepsilon(k)c^\dagger_{\alpha}(k)c_{\alpha}(k)
+v(k)\left[c^\dagger_{\alpha}(k)X_{0\alpha}+
X_{\alpha 0}c_{\alpha}(k)\right]\right\}
+\epsilon_f\sum_{\alpha}X_{\alpha\alpha}.
\end{equation}
The Fermi operator $c^\dagger_{\alpha}(k)$ creates a conduction
band electron with the total angular momentum projection $\alpha$
and the momentum modulus $k$. A rare-earth impurity is described
by the Hubbard operators $X_{ab}$ with algebra
$X_{ab}X_{cd}=\delta_{cb}X_{ad}$, where the index $a=0,\alpha$
enumerates both the nonmagnetic state ($a=0$) and the degenerate
magnetic states ($\alpha=1,\ldots,n$) of the impurity.
The third term is the $f$-level energy of an electron localized on
the impurity, while the second one describes the hybridization of
band and $f$-level electron states with the matrix element $v(k)$.
The Coulomb repulsion on the impurity orbital is assumed to be
very large, so that the multiple occupancy of the impurity $f$-level
is excluded. Within LDA the Anderson model has been diagonalized
by Wiegmann \cite{W} (nondegenerate version) and by Schlottmann
\cite{S} (degenerate version).

It has recently been discerned \cite{RS} that LDA is not necessary
for an exact diagonalization of the basic impurity models of
quantum optics, describing a system of Bose particles with
a nonlinear dispersion coupled to two-level atoms. Making use of some
mathematical analogies between ``magnetic'' and ``optical'' models,
we diagonalize exactly the Hamiltonian (1a) and derive a hierarchy
of the Bethe ansatz equations. The information about the nonlinear
electron dispersion and the energy dependence of the hybridization
is contained only in the nonlinear energy dependence of ``rapidities''
of ``charge'' excitations of the system. If the impurity $f$-level
energy $\epsilon_f$ lies below the conduction band edge, the spectrum
of the system exhibits a novel extremely rich class of charge
excitations with energies lying inside the interband gap around
$\epsilon_f$. These excitations have been excluded obviously
from previous studies by LDA. The gap spectrum is shown to
contain propagating pairs of confined ``gap excitations'', which
do not exist separately from each other, and their heavy bound
complexes. In addition to propagating complexes, we demonstrate
the existence of multielectron gap states localized in the vicinity
of the impurity due to a pinning of a gap bound complex to the impurity.

To diagonalize the model Hamiltonian, it is convenient first to rewrite
Eq. (1a) in terms of electron operators on the ``energy scale'',
$c_\alpha(\epsilon)=(d\varepsilon(k)/dk)^{-1/2}c_{\alpha}(k)$,
with algebra
$\{c_\alpha(\varepsilon),c^\dagger_{\alpha'}(\varepsilon')\}=
2\pi\delta_{\alpha\alpha'}\delta(\varepsilon-\varepsilon')$,
\begin{equation}
H=\sum_{\alpha}\int_{0}^{\infty}\frac{d\varepsilon}{2\pi}
\left\{\varepsilon\,c^\dagger_\alpha(\varepsilon)c_\alpha(\varepsilon)
V(\varepsilon)\left[c^\dagger_\alpha(\varepsilon)X_{0\alpha}+
X_{\alpha 0}c_\alpha(\varepsilon)\right]\right\}
+\epsilon_f\sum_{\alpha}X_{\alpha\alpha},
\end{equation}
where $V(\varepsilon)=(d\varepsilon(k)/dk)^{-1/2}v(k)$.
Note that the energy dependence of the effective coupling $V(\varepsilon)$
is determined now by both the energy dependence of the hybridization
matrix element and the nonlinearity of the electron dispersion.

Now we look for one-particle eigenstates of the model in the form
\end{mathletters}
$$
|\Psi_1\rangle=\sum_{\alpha}A_\alpha\left[{\rm g} X_{\alpha 0}+
\int_{0}^{\infty}\frac{d\varepsilon}{2\pi}V(\epsilon)\phi(\varepsilon)
c^\dagger_\alpha(\varepsilon)\right]|0\rangle,
$$
where $A_\alpha$ is arbitrary, and the vacuum state contains no band
electrons and the impurity is in the nonmagnetic state,
$c_\alpha(\epsilon)|0\rangle=X_{a\alpha}|0\rangle=0$.
The Schr\"odinger equation then reads
\begin{mathletters}
\begin{eqnarray}
&&(-i\partial_\tau-\omega)\phi(\tau|\omega)+{\rm g}(\omega)\delta(\tau)=0,\\
&&(\epsilon_f-\omega){\rm g}(\omega)+\int_{-\infty}^{\infty}d\tau
\Gamma(\tau)\phi(\tau|\omega)=0,
\end{eqnarray}
where $\phi(\tau)$ is the Fourier image of the auxiliary wave function
$\phi(\varepsilon)$, while the effective particle-impurity coupling
$\Gamma(\tau)$ contains the information about the electron dispersion,
\end{mathletters}
$$
\phi(\tau)=\int_{-\infty}^{\infty}\frac{d\varepsilon}{2\pi}
\phi(\varepsilon)e^{i\varepsilon\tau},\;\;\;
\Gamma(\tau)=\int_{0}^{\infty}\frac{d\varepsilon}{2\pi}
V^2(\varepsilon)e^{-i\varepsilon\tau}.
$$
For $\omega>0$, one gets ${\rm g}(\omega)=[h(\omega)+i/2]^{-1}$ and
\begin{equation}
\phi(\tau|\omega)=\frac{h(\omega)-(i/2){\rm sgn}(\tau)}{h(\omega)+i/2}
e^{i\omega\tau},
\end{equation}
where the ``rapidity''
$$
h(\omega)=(\omega+\Sigma'(\omega)-\epsilon_f)/V^2(\omega),
$$
while $\Sigma'(\omega)$ and $\Sigma''(\omega)=V^2(\omega)/2$
are respectively the real and imaginary parts of the self-energy
$\Sigma(\omega)=\int_{0}^{\infty}(d\varepsilon/2\pi)
V^2(\varepsilon)/(\varepsilon-\omega-i0)$.
In what follows we neglect the $\omega$ dependence of $\Sigma'(\omega)$,
$\Sigma'(\omega)\rightarrow\Sigma'(\epsilon_f)$, and regard
$\Sigma'(\epsilon_f)$ as a small correction to the $f$-level
energy due to the hybridization with the band electron states,
$\bar{\epsilon}_f=\epsilon_f-\Sigma'(\omega)$. If $\bar{\epsilon}_f$
lies below the band edge, $\epsilon_f<0$, Eqs. (2) admit also
the discrete mode with the eigenenergy $\omega_d=\bar{\epsilon}_f$,
which obviously corresponds to the electron-impurity bound state.
In the case of a Bose system with gap dispersion, a particle-impurity
bound state has earlier been predicted by John and Wang \cite{JW}.
To keep notations simple, hereafter the bar in $\bar{\epsilon}_f$
is omitted.

The auxiliary wave function (3) is discontinuous, but the electron
wave function
$\psi_\alpha(x|\omega)\equiv\langle c_\alpha(x)|\Psi_1\rangle=
\psi(x|\omega)A_\alpha$ (where the electron operator $c_\alpha(x)$ is
defined as
$c_\alpha(x)=\int_{0}^{\infty}(dk/2\pi)c_{\alpha}(k)\exp{(ikx)}$ \cite{MA})
is continuous and results from the integral ``dressing'' of
the auxiliary function,
$$
\psi(x|\omega)=\int_{-\infty}^{\infty}d\tau u(x;\tau)\phi(\tau|\omega),
$$
with the dressing function
$$
u(x;\tau)=\int_{0}^{\infty}\frac{dk}{2\pi}v(k)
\left(\frac{d\varepsilon(k)}{dk}\right)^{-1}
e^{i(kx-\varepsilon(k)\tau)}.
$$
In LDA, the dressing function is nothing but the delta-function,
$u(x;\tau)\sim\delta(x-\tau)$, and hence the auxiliary wave function
$\phi(\tau|\omega)$ is the same as the electron wave function
$\psi(x|\omega)$.

The idea of auxiliary functions (or ``auxiliary particles'') can
easily be extended to the multielectron case. For instance, the
two-electron wave functions in the energy space are represented as
$\Psi_{\alpha_1\alpha_2}(\varepsilon_1,\varepsilon_2)=
V(\varepsilon_1)V(\varepsilon_2)
\Phi_{\alpha_1\alpha_2}(\varepsilon_1,\varepsilon_2)$ and
$J_{\alpha_1\alpha_2}(\varepsilon)=V(\varepsilon)
G_{\alpha_1\alpha_2}(\varepsilon)$, where the latter describes the
state, in which one of the electrons is localized on the impurity.
In the auxiliary $\tau$-space, the Schr\"odinger equations for the
Fourier images of the auxiliary functions
$\Phi_{\alpha_1\alpha_2}(\tau_1,\tau_2)$ and
$G_{\alpha_1\alpha_2}(\tau)$ are then similar to those within LDA
but with the nonlocal particle-impurity coupling $\Gamma(\tau)$.
It is remarkable that, despite the nonlocal coupling, the two-particle
scattering matrix is still found \cite{RS} in the well-known form:
\begin{equation}
{\bf S}=\frac{h(\omega_1)-h(\omega_2)+i{\bf P}}{h(\omega_1)-h(\omega_2)+i}
\end{equation}
where ${\bf P}=\delta_{\alpha_1\alpha'_2}\delta_{\alpha_2\alpha'_1}$
is the permutation operator. The $S$-matrix satisfies obviously the
Yang-Baxter equations. Hence the multiparticle scattering of the
auxiliary particles is factorized into two-particle ones and the
auxiliary multiparticle wave functions have the ordinary Bethe ansatz
structure. But due to the nonlinear electron dispersion and the nonlocal
electron-impurity coupling, the multielectron wave functions
\begin{equation}
\Psi_{\alpha_1\ldots\alpha_N}(x_1,\ldots,x_N)=\int_{-\infty}^{\infty}
\Phi_{\alpha_1\ldots\alpha_2}(\tau_1,\ldots,\tau_N)
\prod_{j=1}^{N}u(x_j;\tau_j)d\tau_j
\end{equation}
are continuous functions of the coordinates $x_j$. The factorization
of multielectron scattering and the Bethe ansatz construction for
multielectron wave functions are thus {\em hidden} and manifested only
in the limit of large interelectron separations.

Imposing the periodic boundary conditions on the $N$-electron wave
function (5) on the interval of size $L$, we arrive at the following
Bethe ansatz equations (BAE):
\begin{mathletters}
\begin{eqnarray}
&&\exp{(ik_jL)}e_1(\lambda^{(0)}_j)=
\prod_{a_1=1}^{M_1}e_1(\lambda^{(0)}_j-\lambda^{(1)}_{a_1})\\
&&\prod_{\nu=\pm 1}
\prod_{a_{l+\nu}=1}^{M_{l+\nu}}
e_1(\lambda^{(l)}_{a_l}-\lambda^{(l+\nu)}_{a_{l+\nu}})
=-\prod_{b=1}^{M_l}e_2(\lambda^{(l)}_{a_l}-\lambda^{(l)}_{b})
\end{eqnarray}
where $E=\sum_{j}\omega_j$ is the eigenenergy, $k_j\equiv k(\omega_j)$
is the momentum of a charge excitation with the energy $\omega_j$, and
$e_n(x)=(x-in/2)/(x+in/2)$. If $m_\alpha$ is the number of particles
with the color $\alpha$, the numbers $M_l$ are defined by
$M_l=\sum_{\alpha=l}^{n-1}m_\alpha$, $M_0$ being the total number of
electrons, $M_0\equiv N$. It is clear that only charge excitations
with rapidities $\lambda^{(0)}_j\equiv h(\omega_j)$ contain the
information about the electron dispersion and the energy dependence
of the hybridization, while BAE for the ``color'' rapidities
$\{\lambda^{(l)}\}$, $l=1,\ldots,n-1$ coincide with the corresponding
equations in LDA. Therefore, we can use previous results \cite{S}
to classify the solutions of Eqs. (6) as $L\rightarrow\infty$ and
we may focus our studies on charge excitations only.

The charge rapidities can be grouped into a ``string'', which
describe a bound complex of electrons with different colors,
\end{mathletters}
\begin{equation}
h_j=\Lambda^{(l)}+\frac{i}{2}(m+1-2j),
\end{equation}
where $j=1,\ldots,l+1$ enumerates now the string rapidities, and
an arbitrary real $\Lambda^{(l)}$ denotes a rapidity of corresponding
color excitation. Since only electrons of different colors interact,
one can build strings containing up to $m=l+1\leq n$ charge excitations.
Eq. (7) is a solution of BAE, if and only if the imaginary parts of
the rapidity $h_j$ and the corresponding momentum $k_j$ have the same sign,
\begin{equation}
{\rm sgn}\left({\rm Im}\,h_j\right)=
{\rm sgn}\left({\rm Im}\,k_j\right).
\end{equation}
It is easy to understand that solutions of Eq. (8) determine
vicinities on the $\omega$ axis with an attractive interaction
between charge excitations. In the other vicinities, where NC
is not satisfied, the charge excitation coupling is repulsive,
and hence they comprise no strings, except one-particle ones.
Thus, though BAE are valid at arbitrary $\varepsilon(k)$ and $v(k)$,
their solutions are controlled by the properties of analytical
continuations of the functions $k(\omega)$ and $h(\omega)$ in the
complex $\omega$-plane. In LDA, where $k_j=\omega_j$ and
$h_j=(\omega_j-\epsilon_f)/v^2(k_F)$, the necessary condition (NC)
(8) is obviously satisfied at all frequencies. For the sake of
simplicity, we confine ourselves to the case of a weak momentum
dependence of the hybridization matrix element, assuming that
$v(k)$ can be replaced by some constant value $\sqrt{2}v$.
Moreover, we also assume that the electron dispersion has the
simple form $\varepsilon(k)=k^2$, where we set the effective
electron mass equals to $1/2$. Then one obtains
\begin{equation}
k(\omega)=\sqrt{\omega},\;\;\;
h(\omega)=v^{-2}(\omega-\epsilon_f)k(\omega).
\end{equation}

The analytical continuations of the functions $k(\omega)$ and
$h(\omega)$ depend essentially on the position of the real part
of the frequency with respect to the band edge. If the real parts
of all the frequencies $\omega_j$ corresponding to the $h$-string
(7) are positive, BAE with Eq. (9) lead only to some corrections
to the system spectrum found within LDA. For the states of physical
interest near the Fermi energy, these corrections are negligibly small.

But the situation is changed drastically, if the real parts of
all the frequencies $\omega_j$ are assumed to lie below the band edge,
$\omega_j<0$. Let $\omega=\xi+i\eta$, where $\xi<0$, and
$k(\xi\pm i0)=\pm i\kappa(\xi)$, where $\kappa(\xi)=\sqrt{-\xi}$.
To avoid tedious algebraic expressions, we restrict further analysis
to the case $\eta\ll\xi$. Then
$k(\omega)\simeq{\rm sgn}(\eta)[-\eta\kappa'(\xi)+i\kappa(\xi)]$
and $h(\omega)\simeq{\rm sgn}(\eta)[-\eta r'(\xi)+ir(\xi)]$,
where $r(\xi)=v^{-2}(\xi-\epsilon_f)\kappa(\xi)$. Since $\kappa(\xi)$
is positive, NC leads now to the condition $r(\xi)>0$, which
is met only if $\epsilon_f$ is negative and $\xi\in(\epsilon_f,0)$.
Hereafter, we study only gap states lying around $\epsilon_f$,
because they are well separated from the band edge. The remaining
analysis is then simplified by linearizing the function $r(\xi)$
at the point $\xi=\epsilon_f$, $r(\xi)=a(\xi-\epsilon_f)$, where
$a=v^{-2}\kappa(\epsilon_f)$.

Let us start with the simplest case of a two-particle string.
Its complex conjugated rapidities, $h_1=\Lambda^{(1)}+i/2$ and
$h_2=h_1^*=\Lambda^{(1)}-i/2$ are mapped to the corresponding
pair of frequencies, $\omega=\xi+i\eta$ and $\omega^*=\xi-i\eta$,
where $\eta$ is chosen to be positive, $\eta>0$. The pair of
corresponding momenta are given by $k=q+i\kappa(\xi)$ and
$k^*=q-i\kappa(\xi)$, where $q=-\eta\kappa'(\xi)=\eta|\kappa'(\xi)|$,
because $\kappa'(\xi)$ is negative. The real and imaginary
parts of frequencies are easily found as $\xi=\epsilon_f+1/2a$
and $\eta=-\Lambda^{(1)}/a$. Since $\eta>0$, only strings with
negative $\Lambda^{(1)}$ are mapped to the gap states. For the
gap states the real part of rapidities $\Lambda^{(1)}$ determines
the imaginary part of frequencies $\eta$, therefore strings with
different $\Lambda_\beta^{(1)}$ are mapped to pairs with the same
real part $\xi$ but different imaginary parts
$\eta_\beta=\Lambda_\beta^{(1)}/a$.

The expressions obtained describe a confined state of two gap
excitations of the system, because, unlike excitations bound into
ordinary band two-particle complexes, gap excitations do not exist
separately from each other. These results are easily extended
to the case of strings containing an even number of particles,
$m=2\mu$. All pairs of complex conjugated rapidities, $h_j$ and
$h^*_j$, $j=1,\ldots,\mu$, are mapped to the corresponding pairs
of complex conjugated frequencies, $\omega_j=\xi^{(0)}_j+i\eta$ and
$\omega^*_j=\xi^{(0)}_j-i\eta$, and momenta
$k_j=q_j+i\kappa(\xi^{(0)}_j)$ and $k^*_j=q_j-i\kappa(\xi^{(0)}_j)$.
Here $\xi^{(0)}_j=\epsilon_f+(\mu+1/2-j)/a$, $\eta=\Lambda^{(l)}/a$
and $q_j=\eta|\kappa'(\xi^{(0)}_j)|$. The expressions obtained describe
obviously the bound complex of $\mu$ pairs of confined gap excitations
with the eigenenergy per particle
\begin{equation}
E^{(0)}_\mu=2\sum_{j=1}^{\mu}\xi^{(0)}_j=\epsilon_f+\mu/2a.
\end{equation}
Note that the frequencies and momenta of the gap complex do not
have the string structure. The upper index in the above expressions,
$(0)$, indicates that they are derived with linearizing of the
function $r(\xi)$ at the point $\xi=\epsilon_f$. In this
approximation, the energy of a complex is independent of its
momentum per particle,
\begin{equation}
Q=(2\mu)^{-1}\sum_{j=1}^{\mu}q_j\approx(\eta_l/\mu)\sum_{j=1}^{\mu}
|\kappa(\xi^{(0)}_j)|,
\end{equation}
and hence the states obtained are infinitely degenerate. But the
degeneracy is lifted by the first correction to Eq. (10). To find
this correction, one has to keep the next term, $-i(\eta^2/2)r''(\xi)$,
of the Taylor series for the function $h(\omega)$ and the next
term in the expansion of the function $r(\xi)$ at the point
$\xi=\epsilon_f$: $r(\xi)\simeq a(\xi-\epsilon_f)+b(\xi-\epsilon_f)^2$,
where $b=v^{-2}\kappa'(\epsilon_f)<0$. We then find the first order
correction to the energy of a motionless complex and determine the
kinetic energy contribution to the total energy of a complex
in the effective mass approximation,
\begin{equation}
E_\mu=E^{(0)}_\mu-\delta_\mu+Q^2/2m^*_\mu.
\end{equation}
Here $\delta_\mu=\frac{b}{12 a^3}(4\mu^2-1)$ and
$m^*_\mu=\frac{a}{2b\mu^2}\sum_{j}\left(|\kappa'(\xi^{(0)})|\right)^2$
are the band half-width and the effective mass. For small $\mu$,
the bands of propagating complex are very narrow and complexes are
very heavy and even motionless at $Q=0$. But the bandwidth grows
as $\mu^2$, so that large $\mu$ complexes are quite mobile for
large $Q$. At arbitrary $Q$, one needs to study the exact equations
for the parameters of a complex,
${\rm Re}\, h(\xi_j,\eta_j)=\Lambda^{(l)}$ and
${\rm Im}\, h(\xi_j,\eta_j)=(\mu-j+1/2)$.

In the case of an $h$-string containing an odd number of particles,
$m=2\mu+1$, the extra real rapidity is mapped only to the gap state
with $\xi_0=\omega_d=\epsilon_f$, which corresponds to the
electron-impurity bound state. The remaining complex conjugated pairs
of rapidities of an odd string with $\Lambda^{(l)}\rightarrow 0^+$
are then mapped to a deformed motionless gap complex with
$\xi_j=\epsilon_f+(\mu-j+1)$. Since one of the particles of this
complex is bound to the impurity, a complex as a whole is pinned
to the impurity. A pinned complex describes obviously the state
of the system, in which electrons with different colors are localized
in the vicinity of the impurity.

In summary, we have shown that the degenerate Anderson model
is diagonalized exactly by the Bethe ansatz beyond LDA. If the
impurity $f$-level energy lies inside the gap, the spectrum of
the charge excitations of the system exhibits a novel rich class
of gap states, which have obviously been eliminated from previous
studies due to LDA. It should be emphasized in conclusion that
one of the most exciting possible applications of the approach
developed could be an exact treatment of the superconductivity
problem in the presence of magnetic impurities \cite{R}, where
carrier dispersion cannot be linearized around the Fermi level
of a host metal.

I am grateful to S. John for stimulating discussions and to the
Department of Physics at the University of Toronto for kind
hospitality and support during the completion of this work.

\end{document}